\documentclass[11pt]{article}
\usepackage{graphicx}
\usepackage{amssymb}

\newcommand{\be}{\begin{equation}}
\newcommand{\ee}{\end{equation}} 
\newcommand{\lb}{\label}
\newcommand{\ol}{\overline}

\newcommand{\bF}{{\bf f}}
\newcommand{\bg}{{\bf g}}

\newcommand{\bu}{{\bf u}}
\newcommand{\bx}{{\bf x}}

\newcommand{\bsigma}{{\mbox{\boldmath $\sigma$}}}
\newcommand{\bSigma}{{\mbox{\boldmath $\Sigma$}}}
\newcommand{\bomega}{{\mbox{\boldmath $\omega$}}}
\newcommand{\grad}{{\mbox{\boldmath $\nabla$}}}
\newcommand{\bdot}{{\mbox{\boldmath $\cdot$}}}
\newcommand{\btimes}{{\mbox{\boldmath $\times$}}}
\newcommand{\bzed}{{\mbox{\boldmath $0$}}}
\newcommand{\bell}{{\mbox{\boldmath $\ell$}}}
\newcommand{\hn}{\hat{\bf n}}
\newcommand{\hT}{\hat{\bf t}}

\title{Turbulent Flow in Pipes and Channels \\ as Cross-Stream ``Inverse Cascades'' of Vorticity}
\author{Gregory L. Eyink\\{\it {\small Applied Mathematics \& Statistics, 
The Johns Hopkins University, Baltimore, MD 21218}}\\
{\it {\small Center for Nonlinear Studies, Los Alamos National Laboratory,
Los Alamos, NM 87545}}}
\date{ }

\begin{document}

\maketitle
\begin{abstract}
A commonplace view of pressure-driven turbulence in pipes and channels 
is as  ``cascades'' of streamwise momentum toward the viscous layer at 
the wall. We present in this paper an alternative picture of these flows
as  ``inverse cascades'' of spanwise vorticity, in the cross-stream 
direction but away from the viscous sublayer. We show that there is a 
constant spatial flux of spanwise vorticity, due to vorticity conservation,  
and that this flux is necessary to produce pressure-drop and energy 
dissipation.  The vorticity transport is shown to be dominated by viscous 
diffusion at distances closer to the wall than the peak Reynolds 
stress, well into the classical log-layer. The Perry-Chong model 
based on  ``representative'' hairpin/horsehoe vortices predicts a 
single sign of the turbulent vorticity flux over the whole log-layer, whereas 
the actual flux must change sign at the location of the Reynolds-stress
maximum. Sign-reversal may be achieved by assuming a slow power-law 
decay of  the Townsend ``eddy intensity function" for wall-normal distances greater 
than the hairpin length-scale. The vortex-cascade picture presented here has a 
close analogue in the theory of quantum superfluids and superconductors, 
the ``phase slippage'' of quantized vortex lines. Most of our results should 
therefore apply as well to superfluid turbulence in pipes and channels. We 
also discuss issues about drag-reduction from this perspective.  
\end{abstract}

\newpage 

\section{Introduction}\lb{intro}

It is quite standard to view turbulence in pipes, channels and ducts, as well as 
other wall-bounded turbulent boundary layers, as spatial ``cascades'' of momentum
in the wall-normal or cross-stream direction.  For example, we may quote from a 
well-known monograph \cite{TennekesLumley72}:
\begin{quote}
``There exists a close analogy between the spatial structure of turbulent boundary
layers and the spectral structure of turbulence. At sufficiently large Reynolds numbers,
the overall dynamics of turbulent boundary layers is independent of viscosity 
just as the large-scale spectral dynamics of turbulence is. In the wall layer of a 
turbulent boundary layer, viscosity generates a `sink' for momentum, much like the 
dissipative sink for kinetic energy at the small-scale end of the turbulence spectrum.
In particular, the asymptotic rules governing the link between the large-scale 
description and the small-scale description lead to the closely related concepts
of an inertial subrange in the turbulence energy spectrum...and an inertial 
sublayer in wall-bounded shear flows.'' $\,\,\,\,\,\,\,\,\,\,\,\,\,\,\,\,\,\,\,\,\,\,\,\,\,\,\,\,\,\,
\,\,\,\,\,\,\,\,\,\,\,\,\,\,\,\,\,\,\,\,\,\,\,\,\,\,\,\,\,\,\,\,\,\,\,\,\,\,$--- Tennekes \& Lumley (1972)
\end{quote}
In this point of view, the logarithmic mean velocity profile is analogous to the 
Kolmogorov $-5/3$ power-law energy spectrum, and the constant Reynolds 
stress through the inertial sublayer is analogous to the constant energy flux 
through the  inertial subrange. As expressed in the quote above, high momentum
from the outer region of the boundary layer is transported inward by the constant 
Reynolds stress and ultimately transferred to the wall by viscous stress. 

There is, however, another complementary process in wall-bounded turbulence
of at least equal importance. Vorticity generated at the wall is transported  
outward, diffused first by viscosity and subsequently advected by turbulent
velocity fluctuations.  Due to the fundamental hydrodynamic principle of 
vorticity conservation \cite{Helmholtz:1858,Foeppl13}, the wall-normal flux 
of spanwise vorticity is constant over the whole cross-section of the flow. 
The ``sink'' of vorticity at the outer range is provided by cancellation with 
vorticity of the reverse sign transported from the wall on the opposite side 
of the flow. There is thus a spatial ``inverse cascade'' of vorticity co-existing 
with the ``forward cascade'' of momentum. Furthermore, it may be shown that 
the constant flux of spanwise vorticity in this ``inverse cascade'' is crucially 
related to the energy dissipation and wall drag in turbulent pipe and channel 
flows. This relation was first observed in early work of Taylor \cite{Taylor32}, 
but seems not to have been as widely appreciated as it deserves in the fluid 
dynamics community. Although there have been very many studies of the Reynolds 
stress---or turbulent momentum transport---in pipe and channel flow, there 
have been far fewer investigations of the turbulent vorticity-transport tensor in 
classical fluid turbulence. Notable exceptions are the experimental works by Klewicki 
and collaborators \cite{Klewicki89,Klewickietal94,Priyadarshanaetal07,Klewickietal07}
and a theoretical and DNS study of Bernard \cite{Bernard90}.  In the meantime, 
Taylor's observation of the relation between cross-stream transport 
of vorticity and downstream pressure drop has been rediscovered 
in the field of quantum superfluids, where it goes by the name of the 
``Josephson-Anderson relation'' \cite{Anderson66,Huggins70}. The cross-stream 
motion or Òphase slippageÓ of quantized vortex lines is widely recognized to be 
a key mechanism of energy dissipation in quantum superfluids and 
superconductors \cite{Zimmerman93,Packard98}. 

The purpose of this paper is to expound in detail the picture of pipe 
and channel flows, or other pressure-driven turbulent flows, as 
spatial ``inverse cascades'' of vorticity co-existing with the more 
commonly recognized ``forward cascade'' of momentum. Although 
several of the results presented here are to be found in the existing 
literature, there seems to be no systematic presentation of this fundamental 
point of view of wall-bounded turbulence. It is hoped that the discussion 
here will help to stimulate some further experimental and numerical efforts 
to elucidate the relevant vortex dynamics in turbulent flows. In addition, 
this paper will discuss the close analogy between vortex dynamics in 
classical turbulence and in quantum superfluids. A very interesting paper 
on this subject has been written by Huggins \cite{Huggins94}, from the 
vantage of the low-temperature physicist. We shall further elaborate upon 
his discussion, filling in some of the details on classical fluid turbulence. 
Needless to say,  comparisons between such disparate areas of physics 
can often be extremely illuminating and can stimulate new questions and 
different points of view in both subjects. 

\newpage

\section{Vorticity Conservation and Generation at the Boundary}\lb{vorticity}

We begin by briefly reviewing some basic results on vorticity in classical 
fluid dynamics, its associated conservation law and its generation at 
solid walls. This is necessary in order to set the stage for our following 
discussion on wall-bounded turbulence. Furthermore, even this very 
classical subject is not without controversy, as there is currently some 
disagreement about the proper form of the vorticity source at the wall
\cite{Lyman90,WuWu96}. 

\subsection{The Helmholtz Equation as Conservation Law}

The evolution of vorticity in classical incompressible fluids is described by 
the equation of  Helmholtz \cite{Helmholtz:1858}
\be \partial_t \bomega + (\bu\bdot\grad)\bomega = 
(\bomega\bdot\grad)\bu+ \nu\bigtriangleup \bomega+\grad\btimes \bF, 
\lb{Helm} \ee
supplemented by terms representing viscous diffusion with kinematic 
viscosity $\nu$ and the curl of a non-potential body force 
$\bF$ (e.g. from electromagnetic stirring \cite{CrawfordKarniadakis97} or stress 
from polymer additives \cite{Dubiefetal05,Kimetal08}). Because this equation is 
obtained by taking the curl of the momentum equation, or incompressible Navier-Stokes 
equation, it must take the form of a local conservation law. Explicitly, (\ref{Helm}) 
can be rewritten as  \cite{Huggins94}         
\be    \partial_t \bomega + \grad\bdot \bSigma = \bzed \lb{vort-cons} \ee
or $\partial_t\omega_j + \partial_i\Sigma_{ij}=0,$ where
\be    \Sigma_{ij}=u_i\omega_j-u_j\omega_i
+\nu\left(\frac{\partial\omega_i}{\partial x_j}- \frac{\partial\omega_j}{\partial x_i}\right) 
+ \epsilon_{ijk} f_k. \lb{Sigma}  \ee   
This anti-symmetric tensor represents spatial flux of the $j$th component of vorticity 
in the $i$th coordinate direction. It plays a fundamental role in all of our following 
considerations. The first term on the righthand side of (\ref{Sigma}) represents 
convective transport, while the second term represents spatial transport by 
vortex-shearing, the third term describes viscous diffusion and the fourth term
gives the induced motion of vorticity due to Magnus effect of the body force. 

Because vorticity arises as the curl of the velocity, $\bomega=\grad\btimes\bu,$ 
the above conservation law must be in some sense ``trivial''. This is the content of
F\"{o}ppl's Theorem \cite{Foeppl13}, which states that the total vorticity integrated 
over the flow domain $\Omega$ must vanish:
$\int_\Omega \bomega d^3 x = \bzed.$
This result holds in an infinite domain with sufficiently rapidly decaying vorticity 
at infinity, or in a bounded domain with uniformly moving, solid walls (but 
not, for example, in a rotating container). The standard proof is based on the 
simple identity $\bomega=\grad\bdot (\bomega\bx),$ which implies by the 
divergence theorem that 
$\int_\Omega \bomega d^3 x = \int_{\partial \Omega} \bx (\bomega \bdot \hn) dA $
for the outward-pointing normal $\hn$ on the bounding surface $\partial\Omega.$ 
On a solid wall in uniform motion, $\bomega \bdot \hn=0,$ because of stick boundary 
conditions for the velocity field. There is also no contribution from boundary points 
at infinity, if $|\bomega|=o(|\bx|^{-3})$ as $|\bx|\rightarrow\infty.$ This result shows 
that complete cancellation of vorticity must occur when integrated over the entire
domain of such flows, despite strong localized vortex structures of various orientations 
distributed throughout  the fluid.

 \subsection{Lighthill's Theory of Vorticity Generation at Solid Walls}

If the vorticity conservation law (\ref{vort-cons}) is integrated over the flow domain, 
one obtains 
\be  \frac{d}{dt} \int_\Omega \bomega d^3 x = \int_{\partial \Omega} \bSigma\hn \,dA. 
        \lb{vort-cons-global} \ee
Of course, under the conditions of F\"{o}ppl's Theorem, which we hereafter assume, 
$ \int_{\partial \Omega} \bSigma\hn \,dA=\bzed.$ As argued by Lyman \cite{Lyman90}, 
the relation (\ref{vort-cons-global}) motivates one to consider $\bsigma=\bSigma\hn$ 
as a source density for vorticity at the wall, with strength $|\bsigma|$ in the direction 
$\hat{\bsigma}=\bsigma/|\bsigma|.$ As a consequence of anti-symmetry of $\bSigma,$
the vortex lines generated are parallel to the surface, since $\bsigma\bdot\hn=0$
\cite{footnote1}.  However, Wu and Wu \cite{WuWu96} have pointed out that 
such a global argument is insufficient to identify $\bsigma$ as a local source of vorticity, 
since it may be altered by adding any perturbation $\delta \bsigma$ such that 
$\int_{\partial\Omega} \delta\bsigma\,dA=\bzed.$  In the Appendix A, we present 
a proof based upon the Kelvin Theorem which shows that $\bsigma=\bSigma\hn$
does represent a localized source of vorticity at the boundary. This holds in the 
following precise sense: If one considers {\it any} loop $C$ in the fluid with part of 
its length lying in the boundary $\partial \Omega$, then the contribution of the boundary 
segment to the time-derivative of the vorticity flux through $C$ is given by
$$ \int_{C \cap \partial \Omega} \bsigma \cdot (\hn \btimes \hT) ds $$
where $\hT$ is the unit tangent vector along the curve $C$, for a specified 
orientation. With this convention the flux is measured positive in the direction 
to the righthand side of the curve $C$ as one moves along it, i.e. parallel to 
$\hn \btimes \hT.$

Using the expression (\ref{Sigma}) for $\bSigma,$ it follows that 
\be \bsigma=-\hn\btimes \nu(\grad\btimes\bomega) + \hn\btimes\bF \ee
due to the no-through condition $\bu\bdot\hn=0$ for an impermeable wall and 
the result $\bomega\bdot\hn=0$ for a solid wall in uniform motion. Further insight 
and simplification can be obtained if one uses the Navier-Stokes momentum balance 
$D_t \bu = - \grad p - \nu \grad \btimes \bomega +\bF$ to give \cite{footnote2}
$$ \bsigma = \hn \btimes (\grad p + D_t \bu). $$
This form of the vorticity source density was first derived by Lighthill \cite{Lighthill63} and 
Morton \cite{Morton84} who obtained the terms from the tangential pressure gradient and 
the tangential boundary acceleration, respectively. The term identified by Morton, $\hn
\btimes (D_t \bu)$, contributes only if the boundary is accelerating continuously or impulsively. 
If we restrict our attention to constant-velocity (or stationary) boundaries, then we recover the 
original Lighthill (1963) result
\be  \bsigma = \hn \btimes (\grad p). \lb{lighthill} \ee
This formula has several remarkable implications \cite{Morton84}. First, 
one can see that the generation of vorticity is essentially {\it inviscid}.
Second, the generated vortex lines on boundary $\partial \Omega$ are parallel to the 
surface isobars,  or lines of constant pressure $p$.

\subsection{Mean Flux of Vorticity in Pressure-Driven Flows }\lb{vort-flux-press-grad}

In turbulent flow, the vorticity balance (\ref{vort-cons}) may be considered 
in Reynolds-averaged form:
\be    \partial_t \ol{\bomega} + \grad\bdot \ol{\bSigma} = \bzed \ee
where the mean vorticity flux tensor is
\be    \ol{\Sigma}_{ij}=\ol{u}_i\ol{\omega}_j-\ol{u}_j\ol{\omega}_i 
                       + \ol{u_i'\omega_j'-u_j'\omega_i'} 
+\nu\left(\frac{\partial\ol{\omega}_i}{\partial x_j}
- \frac{\partial\ol{\omega}_j}{\partial x_i}\right) + \epsilon_{ijk} \ol{f}_k. 
\lb{Sigma-tensor} \ee
The first term $\ol{u}_i\ol{\omega}_j-\ol{u}_j\ol{\omega}_i$ represents the 
transport of mean vorticity by the mean flow. On the other hand, the term
$\ol{u_i'\omega_j'-u_j'\omega_i'}$ represents the transport of mean vorticity 
by the fluctuations and is the analogue of the Reynolds stress in the 
ensemble-averaged momentum balance.   

We now derive a crucial relationship between mean vorticity transport and mean
pressure gradients. The Navier-Stokes equation may be written as 
$ \partial_t u_k = \frac{1}{2}\epsilon_{klm}\Sigma_{lm}-\partial_k \left(p+U
+\frac{1}{2}|\bu|^2\right)$
in terms of $\bSigma$ and the potential $U$ of any conservative body force 
(e.g. gravity). In statistically stationary turbulence, the time-derivative averages to zero,
so that one obtains
\be \partial_k(\ol{p+U+\frac{1}{2}|\bu|^2}) = \frac{1}{2}\epsilon_{klm}\ol{\Sigma}_{lm},\,\,\,\,\,\,
       \ol{\Sigma}_{ij}=\epsilon_{ijk}\partial_k(\ol{p+U+\frac{1}{2}|\bu|^2}). \lb{classical-JA} \ee
To our knowledge, this relation was first derived by Taylor \cite{Taylor32} for a 
general turbulent shear-flow, assuming strict two-dimensionality of the fluctuations.
His  work is discussed in Section \ref{channel} below, on turbulence
in pipes and channels. More than thirty years later, this result was rediscovered
by Anderson \cite{Anderson66}, who considered inviscid flow with no body 
forces. His work will be considered in Section \ref{superfluid} on quantum 
superfluids. The full relationship (\ref{classical-JA}) was derived by Huggins 
\cite{Huggins94}, as a  ``Josephson-Anderson relation''  for classical turbulence.     
          
Another useful form of this relation can be derived by using the standard 
vector-calculus identity 
$(\ol{\bu}\bdot\grad)\ol{\bu}=\ol{\bu}\btimes\ol{\bomega}-\grad(\frac{1}{2}|\ol{\bu}|^2)$
to write the Reynolds-averaged Navier-Stokes equation as            
$$ \ol{D}_t \ol{u}_k = \frac{1}{2}\epsilon_{klm}\ol{\Sigma}_{lm}^*-\partial_k \ol{p}^*,$$
with 
\be    \ol{\Sigma}_{ij}^*=\ol{u_i'\omega_j'-u_j'\omega_i'} 
+\nu\left(\frac{\partial\ol{\omega}_i}{\partial x_j}
- \frac{\partial\ol{\omega}_j}{\partial x_i}\right) + \epsilon_{ijk} \ol{f}_k 
\lb{Sigma-star-tensor} \ee
and $\ol{p}^*=\ol{p}+U+k,$ where $k=\frac{1}{2}\ol{|\bu'|^2}$ 
is the turbulent kinetic energy
and $\ol{D}_t=\partial_t+\ol{\bu}\bdot\grad$ is the material time-derivative for the mean
flow. Thus, in cases where $\ol{D}_t\ol{\bu}=\bzed,$
\be \partial_k\ol{p}^* = \frac{1}{2}\epsilon_{klm}\ol{\Sigma}_{lm}^*,\,\,\,\,\,\,
       \ol{\Sigma}_{ij}^*=\epsilon_{ijk}\partial_k\ol{p}^*. \lb{classical-JA*} \ee
In this version of the relationship, the vorticity transport by the mean flow is omitted
and gradients in the ``turbulent enthalpy'' $\ol{p}^*=\ol{p}+U+k$ are considered. 

We now consider the implications of these results for turbulent channel and pipe flows.

\section{Channel and Pipe Flows}\lb{channel}

As usual, channel flow refers to fluid motion between two parallel, infinite plane walls 
separated by distance $2h,$ with a pressure gradient parallel to the planes. Our notations
follow those of \cite{TennekesLumley72}. Thus, we take $x$ to be the streamwise 
direction (along the pressure gradient), $z$ the spanwise direction, and $y$ the 
wall-normal or cross-stream direction, with $y=0$ at the bottom plate. The velocity 
field is then represented as $\bu=(u,v,w)$ in terms of its streamwise, wall-normal, 
and spanwise components. It is assumed that the turbulence is fully developed, 
in the sense that it is statistically stationary and that all averages are independent 
of $z$ and also (except for the mean pressure $\ol{p})$ independent of $x$. 
The only non-vanishing component of mean velocity is $\ol{u}(y),$ with a maximum
value $\ol{u}_c=\ol{u}(h)$ at the channel centerplane. The mean vorticity has only 
a spanwise component, $\ol{\omega}_z(y)=-\partial\ol{u}/\partial y,$ which is negative
for $y<h$ and positive for $y>h.$

Pipe flow refers instead to fluid motion down an infinitely long, cylindrical pipe with 
circular cross section of radius $\rho,$ driven by a pressure gradient down the pipe axis.
We use standard cylindrical coordinates $(r,\theta,z)$, so that $z$ represents the 
streamwise/axial direction, $\theta$ the azimuthal direction, and $r$ the 
radial/wall-normal/cross-stream direction. We write the velocity as $\bu=
(u_r,u_\theta,u_z)$ and similarly for other vector quantities.  In fully developed 
pipe turbulence, all average quantities are independent of $t, \theta$ and $z,$
except that mean pressure $\ol{p}$ decreases linearly with $z$ The only non-vanishing 
component of mean velocity is $\ol{u}_z(r),$ with a maximum value $\ol{u}_c=\ol{u}_z(0)$ 
at the center axis. The mean vorticity has only an azimuthal component, 
$\ol{\omega}_\theta(r)=-\partial\ol{u}_z/\partial r>0.$ Because these two 
problems are so similar, we shall give full details only for channel flow and then 
comment upon the relevant differences for pipe flow. 

\subsection{Mean Vorticity Flux and Energy Dissipation}\lb{vorticity-flux}

In channel flow, the relations (\ref{classical-JA*}) become
\be \overline{v'\omega_z'-w'\omega_y'}-\nu\frac{\partial\ol{\omega}_z}{\partial y}
                                                              =\ol{\Sigma}_{yz}^*=\frac{\partial \ol{p}^*}{\partial x} 
                                                              \lb{span-vort-flux} \ee
\be \overline{w'\omega_x'-u'\omega_z'}
                                                               =\ol{\Sigma}_{zx}^*=\frac{\partial \ol{p}^*}{\partial y} 
                                                               \lb{stream-vort-flux} \ee
\be   \overline{u'\omega_y'-v'\omega_x'}
                                                               =\ol{\Sigma}_{xy}^*=0. \lb{norm-vort-flux} \ee   
Note that $\ol{\Sigma}_{ij}=\ol{\Sigma}_{ij}^*$ for all $i,j$ except  $\ol{\Sigma}_{zx}=
-\ol{\Sigma}_{xz}=-\ol{u}\,\ol{\omega}_z+\ol{\Sigma}_{zx}^*.$ As should be clear 
from the general derivation, these relations are just the usual momentum balance
equations written in a different form. For example, it is easy to check that 
\be \overline{v'\omega_z'-w'\omega_y'}=-\frac{\partial}{\partial y}\ol{u'v'},\,\,\,\,\,\,
      -\nu\frac{\partial\ol{\omega}_z}{\partial y}=\nu\frac{\partial^2\ol{u}}{\partial y^2} 
      \lb{vortflux-turbforce} \ee
See also \cite{TennekesLumley72}, eq.(3.3.13), or \cite{Klewicki89}.  However, now 
(\ref{span-vort-flux})-(\ref{norm-vort-flux}) are seen to have another, equally important 
interpretation: they relate the pressure gradients in various directions to 
the corresponding cross-fluxes of a transverse component of vorticity. The first
relation (\ref{span-vort-flux}) was already derived by Taylor \cite{Taylor32} 
in a slightly modified form for two-dimensional flow, as mentioned earlier.
Taylor employed this relation, which is exact, to motivate his well-known 
``vorticity transfer hypothesis,'' which assumes gradient-transport of vorticity. 
As discussed in \cite{TennekesLumley72}, section 3.3, the latter hypothesis 
ignores vortex-stretching interactions and has limited validity (but see \cite{Bernard90} 
for a refinement). As a consequence, this paper of Taylor's has fallen into a bit of 
disrepute and it is not usually noted that a significant exact result was derived there.
                                                          
Of the relations (\ref{span-vort-flux})-(\ref{norm-vort-flux}), the most 
important is indeed (\ref{span-vort-flux}),  since it relates the cross-stream 
flow of spanwise vorticity $\ol{\Sigma}_{yz}$ to the streamwise pressure 
gradient $\partial\ol{p}/\partial x$. Note that  $\partial \ol{p}^*/\partial x=
\partial\ol{p}/\partial x,$ since $U$ is zero and $k$ is 
independent of $x$. It is well-known that  the 
mean streamwise pressure-gradient is constant across a cross-section of the channel, 
i.e. $\partial\ol{p}/\partial x=-\gamma_*$ independent of $y.$ However, in standard 
derivations (e.g. \cite{TennekesLumley72}, section 5.2) this appears to be an 
``accident''. Here it is seen that the fundamental reason is the conservation of vorticity:
$$ \frac{\partial}{\partial y} \left(\frac{\partial \ol{p}}{\partial x}\right)=\partial_y\ol{\Sigma}_{yz}=0.$$
It is easy to determine its constant value 
by the balance of stress at the two walls against the pressure-gradient applied over a 
channel cross-section, giving $-2 u_*^2=(2h)\frac{\partial \ol{p}}{\partial x},$ or 
$\frac{\partial \ol{p}}{\partial x}=-\gamma_*=-u_*^2/h$ in terms of the friction 
velocity $u_*.$  

The physical picture that emerges
is very simple. In view of Lighthill's relation (\ref{lighthill}), mean spanwise vorticity 
$\ol{\omega}_z$ is generated at the two walls, at a rate given by $\partial \ol{p}/\partial x
=-\gamma_*$. Due to the change in the direction of outward normal vector $\hn,$ this 
vorticity is negative at $y=0$ and positive at $y=2h.$ The negative vorticity at the bottom 
wall is transported upward and the positive vorticity at the upper wall is transported 
downward, leading to a constant negative mean vorticity flux $\ol{\Sigma}_{yz}=
-\gamma_*.$ Although the actual geometry and dynamics of vortex lines is quite 
complex (see subsections \ref{scaling-structure}-\ref{exp-dns} below), what happens 
``on average'' is that straight vortex lines are generated in the spanwise direction 
of equal strengths of vorticity, negative at the bottom and positive at the top. These 
then ``cascade'' toward the channel centerplane, where they annihilate each other.

It is interesting to compare the above with the more conventional picture of channel flow 
as a ``momentum cascade'' \cite{TennekesLumley72}. Integrating the mean
$x$-momentum balance once in $y$ gives a standard formula for the total stress, 
Reynolds stress plus viscous stress:
\be  \ol{T}_{xy}=\ol{u'v'}-\nu\frac{\partial \ol{u}}{\partial y} = 
                                                    -u_*^2 \left(1-\frac{y}{h}\right). \lb{stress} \ee
For $0<y\ll h,$ it follows that $\ol{T}_{xy}\approx -u_*^2$ (and likewise $\ol{T}_{xy}\approx 
u_*^2$ for $0<2h-y\ll h$), leading to the notion of a constant momentum flux range.  
In fact, however, $\ol{\Sigma}_{yz}=-\partial_y\ol{T}_{xy}=-u_*^2/h,$ so that the stress can 
never be strictly constant in any range of $y,$ whereas the vorticity flux is exactly 
constant $\ol{\Sigma}_{yz}=-\gamma_*$ over the whole channel cross-section.   

The importance of the cross-stream flux of spanwise vorticity is that it is directly  
related to energy dissipation. By considering the balance equations of energy in the 
mean flow $(1/2)|\ol{u}|^2$ and energy in the fluctuations $k=(1/2)\ol{|\bu'|^2}$, 
it is easy to show that 
\be \ol{u}_b \left(-\frac{\partial \ol{p}}{\partial x}\right)
=\frac{1}{2h}\int_{0}^{2h} \left[\varepsilon(y)+\nu\left(\frac{\partial \ol{u}}{\partial y}\right)^2\right] \,dy
= \varepsilon_b^{{\rm tot}}. \lb{dissipation}  \ee
Here $ \ol{u}_b= \frac{1}{2h}\int_{0}^{2h} \ol{u}(y)\,dy$ is the bulk velocity, averaged 
over a cross-section of the channel, and $\varepsilon(y)=\nu\ol{|\grad\bu^\prime |^2}$
is the ``turbulent dissipation'' (per unit mass) in the velocity fluctuations. To derive 
(\ref{dissipation}) we have used the result that turbulence production balances turbulent 
dissipation when integrated over the channel width:
$$ \frac{1}{2h}\int_{0}^{2h} \left[-\ol{u'v'}\frac{\partial \ol{u}}{\partial y}\right]\, dy 
    =\frac{1}{2h}\int_{0}^{2h} \varepsilon(y)\, dy.
$$
The relation (\ref{dissipation}) can be rewritten by multiplication with the mass 
density $\varrho$ as 
\be  \ol{J}_b (-\ol{\Sigma}_{yz}) = \varrho  \varepsilon_b^{{\rm tot}} \lb{JA-diss} \ee
where $\ol{J}_b=\varrho \ol{u}_b$ is the mean mass flux through the pipe and 
$\varrho  \varepsilon_b^{{\rm tot}}$ is the bulk-average energy dissipation 
per unit volume. This relation shows the direct connection of vorticity flux
with energy dissipation. As we shall see in Section \ref{superfluid}, (\ref{JA-diss})
is a special case of a far more general relation, valid for both classical and 
quantum fluids. The practical significance of (\ref{span-vort-flux}), i.e. $\ol{\Sigma}_{yz}=
\frac{\partial \ol{p}}{\partial x}$, and (\ref{JA-diss}) is that both pressure-drop
and energy dissipation can be reduced in turbulent flow by decreasing the 
cross-stream flux of spanwise vorticity. Any mechanism that interrupts the 
``vorticity cascade'' will lead to turbulent drag reduction. Body-forces from 
electromagnetic forcing \cite{CrawfordKarniadakis97} and polymer additives
\cite{Dubiefetal05,Kimetal08} are well-known devices.  

In pipe flow, the relations (\ref{classical-JA*}) become
\be \overline{u_r'\omega_\theta'-u_\theta'\omega_r'}
       -\frac{\nu}{r}\frac{\partial}{\partial r}(r\ol{\omega}_\theta)
                                                              =\ol{\Sigma}_{r\theta}^*=\frac{\partial \ol{p}^*}{\partial z} 
                                                              \lb{azimuth-vort-flux} \ee
\be \overline{u_\theta'\omega_z'-u_z'\omega_\theta'}
                                                               =\ol{\Sigma}_{\theta z}^*=\frac{\partial \ol{p}^*}{\partial r} 
                                                               \lb{axial-vort-flux} \ee
\be   \overline{u_z'\omega_r'-u_r'\omega_z'}
                                                               =\ol{\Sigma}_{z r}^*=0. \lb{radial-vort-flux} \ee   
Similarly as before, $\ol{\Sigma}_{ij}=\ol{\Sigma}_{ij}^*$ for all $i,j$ except  $\ol{\Sigma}_{\theta z}=
-\ol{\Sigma}_{z\theta}=-\ol{u}_z\,\ol{\omega}_\theta+\ol{\Sigma}_{\theta z}^*$ and  
$\frac{\partial \ol{p}^*}{\partial z}=\frac{\partial \ol{p}}{\partial z}$. The relation 
(\ref{azimuth-vort-flux}), or  $\ol{\Sigma}_{r\theta}=\frac{\partial \ol{p}}{\partial z},$
is the crucial one.  The radial-independence of the streamwise pressure-gradient 
follows from vorticity conservation:
$$ \frac{\partial}{\partial r} \left(\frac{\partial \ol{p}}{\partial z}\right)
                         =\partial_r\ol{\Sigma}_{r\theta}=0.$$
The constant value $\frac{\partial \ol{p}}{\partial z}=-\gamma_*$ can be inferred
from balancing the stress at the wall  with the pressure-gradient applied across
a cross-section of the pipe, i.e. $-2\pi r\cdot u_*^2=\pi r^2\cdot \frac{\partial \ol{p}}{\partial z}$
so that $\frac{\partial \ol{p}}{\partial z}=-\gamma_*=-2u_*^2/r.$  The cascade picture 
``on average'' is of vortex rings of positive azimuthal vorticity generated at the 
pipe wall and collapsing radially inward to annihilate at the axis.                      

\subsection{Scaling Theories and Vortex Structures}\lb{scaling-structure}

The previous considerations have all been exact. We now develop additional 
insight using theoretical assumptions, involving scaling hypotheses and 
structural modeling.  We consider explicitly only channel flow although 
the discussion applies as well to pipe flow. 

\subsubsection{Reynolds-Number Scaling Theories}

We first examine the consequences of high-Reynolds number asymptotics. 
We shall discuss the classical asymptotic matching theory that leads to the 
logarithmic law of the wall, particularly as laid out in the review of  Panton 
\cite{Panton05}. Many of the results that we deduce below from this theory can 
be obtained under more general assumptions, e.g. see \cite{Weietal05,Fifeetal05}.
The classical theory has, however, the advantages of simplicity and familiarity,
so that we shall restrict ourselves to it here. As usual, one introduces an 
inner scaling $y^+=u_*y/\nu,$ $\ol{u}^+=\ol{u}/u_*=f(y^+,Re_*),$
$-{\ol{u'v'}}^+=-\ol{u'v'}/u_*^2=g(y^+,Re_*)$ with $Re_*=\frac{u_*h}{\nu}=h^+.$
The stress relation (\ref{stress}) then becomes
$$ g(y^+,Re_*) + \frac{\partial f}{\partial y^+}(y^+,Re_*) = 1-\frac{y^+}{Re_*}. $$
The standard scaling hypothesis of the ``law of the wall'' is that  
$f(y^+,Re_*)\sim f_0(y^+)$ and  $g(y^+,Re_*)\sim g_0(y^+)$ for $Re_*\gg 1,$
and, in that case, $ g_0(y^+)+ f'_0(y^+) = 1. $ Matching to an outer law 
leads to the prediction of a ``logarithmic layer'' with 
$$ f_0(y^+) \sim \frac{1}{\kappa}\ln y^+ +B,\,\,\,\,g_0(y^+)\sim 1-\frac{1}{\kappa y^+} $$
for $h^+\gg y^+\gg 1.$ The above results are asymptotically valid only in the 
inner layer, with $y\ll h.$ However, as discussed by Panton \cite{Panton05}, 
these inner laws can be combined with outer laws in a standard technique 
of ``composite expansions'' to yield high-$Re_*$ asymptotic formulae that are 
uniformly valid over the whole channel width. In particular, a composite 
expansion for the Reynolds stress is
\be  {-\ol{u'v'}}^+ \simeq g_0(y^+) -\frac{y^+}{Re_*} \lb{composite} \ee
As shown in \cite{Panton05}, Fig.40, (\ref{composite}) is a very accurate 
approximation, even at rather low values of $Re_*.$ This leads to an important
consequence. Taking the derivative of ${-\ol{u'v'}}^+ \simeq 1-\frac{1}{\kappa y^+}
-\frac{y^+}{Re_*},$ one finds \cite{Panton05} 
\be \frac{d}{dy^+}({-\ol{u'v'}}^+) = \frac{1}{\kappa y^{+\,2}}-\frac{1}{Re_*}, 
       \lb{wall-scaled-force} \ee
so that the peak of the Reynolds stress appears within the logarithmic layer at 
$y_p^+ \sim \sqrt{\frac{Re_*}{\kappa}}$
with maximum value ${-\ol{u'v'}}^+_p \sim 1- \frac{2}{\sqrt{\kappa Re_*}}.$
This scaling of the location of the peak Reynolds stress has been well-verified 
empirically \cite{LongChen81,Sreenivasan87,SreenivasanSahay97}.

There are immediate implications for the turbulent vorticity flux, since, by the 
first line of (\ref{vortflux-turbforce}), it is the derivative of the Reynolds stress:
$\frac{\partial}{\partial y}(-\ol{u'v'})=\overline{v'\omega_z'-w'\omega_y'}.$ 
It follows that 
\be  \begin{array}{l}
        \mbox{$\overline{v'\omega_z'-w'\omega_y'}>0$ for $y^+<y^+_p$}  \cr
        \mbox{$\overline{v'\omega_z'-w'\omega_y'}<0$ for $y^+>y^+_p.$}  \cr 
                                                                       \end{array} \lb{meso} \ee
The condition of constant cross-stream vorticity flux which we derived in 
section \ref{channel} states that
$$ \overline{v'\omega_z'-w'\omega_y'} -\nu\frac{\partial\ol{\omega}_z}{\partial y}=
                 -\frac{u_*^2}{h}<0.$$
The viscous term is negative over the whole channel width, since mean 
spanwise vorticity is negative at $y=0$ and increasing monotonically to 
zero at $y=h,$
so that   $\partial\ol{\omega}_z/\partial y>0$ 
for all $y.$ However, the turbulent transport term is negative only for $y^+>y_p^+.$
These facts have been noted before \cite{Weietal05,Fifeetal05}. We restate them 
here in order to emphasize that {\it mean cross-stream transport of spanwise 
vorticity is dominated  for $y^+<y^+_p$ by viscous diffusion,} which gives the 
contribution of largest magnitude. This result does not contradict any 
of the assumptions in the derivation of the log-layer---and indeed is a consequence 
of that derivation---but it does contradict a common belief that ``the overall dynamics 
of turbulent boundary layers is independent of viscosity'' \cite{TennekesLumley72}. 

Further insight can be 
obtained by considering the constant-flux condition in inner-scaling:
$$  \overline{v'\omega_z'-w'\omega_y'}^+  - \frac{\partial\ol{\omega}_z^+}{\partial y^+}
           = -\frac{1}{Re_*}, $$
with 
$$ \overline{v'\omega_z'-w'\omega_y'}^+= g'(y^+,Re_*), \,\,\,\,\,\,
      \partial\ol{\omega}_z^+/\partial y^+ = -f''(y^+,Re_*). $$
Right at the wall,  $g'(0,Re_*)=-1/Re_*$ and $f''(0,Re_*)=0,$ so that 
the vorticity transport is entirely due to viscous diffusion in close 
vicinity of the wall.  In the range $1\ll y^+<y^+_p$ the two functions
$g',\,\,f''$ are of order unity, so that the small difference $-1/Re_*$
must be due to near-cancellation of  almost equal quantities.  Thus,
the viscous and turbulent transport of vorticity are both relatively large
and the negative residual value of the cross-stream flux is due to a slight 
dominance of the viscous transport. It is only for $y^+\gtrsim y^+_p$  
that the turbulent transport dominates and that one can properly 
think of an ``inertial sublayer'' dominated by nonlinear interactions.
For more discussion of these points, see \cite{Weietal05,Fifeetal05}.

One of the consequences of these considerations is that the 
mean velocity profile $\bar{u}(y)$ plays a critical role in the vorticity cascade 
over the range $y<y_p,$ since the viscous flux equals $\nu d^2\bar{u}(y)/dy^2.$
Conversely, the statistics of vorticity production and transport at the boundary
are a crucial factor in determining the mean profile; for instance, see 
\cite{Dubrulleetal01}. This may have significant implications on understanding 
the mechanism of polymer drag-reduction \cite{Dubiefetal05,Kimetal08}, for example.
As can be seen from the general expression (\ref{Sigma-star-tensor}), mean 
vorticity flux for turbulent pipe flow with a polymer additive will consist of three 
distinct contributions:  the nonlinear turbulent transport, viscous transport, 
and direct transport by Magnus effect of polymeric forces. Given the known 
drag-reduction effects of polymers, it follows from our discussion that wall-normal 
vorticity flux must also be drastically reduced by polymer additives. However, 
this may occur {\it a priori} in several possible ways. For example, for $y<y_p$
either the viscous transport may decrease in magnitude or the turbulent
transport (which is of opposite sign) may increase in magnitude. Further 
research will be necessary to see which of these possibilities is realized.

\subsubsection{Vortex Structure Models}

\begin{figure}
\centerline{\includegraphics[width=300pt,height=150pt]{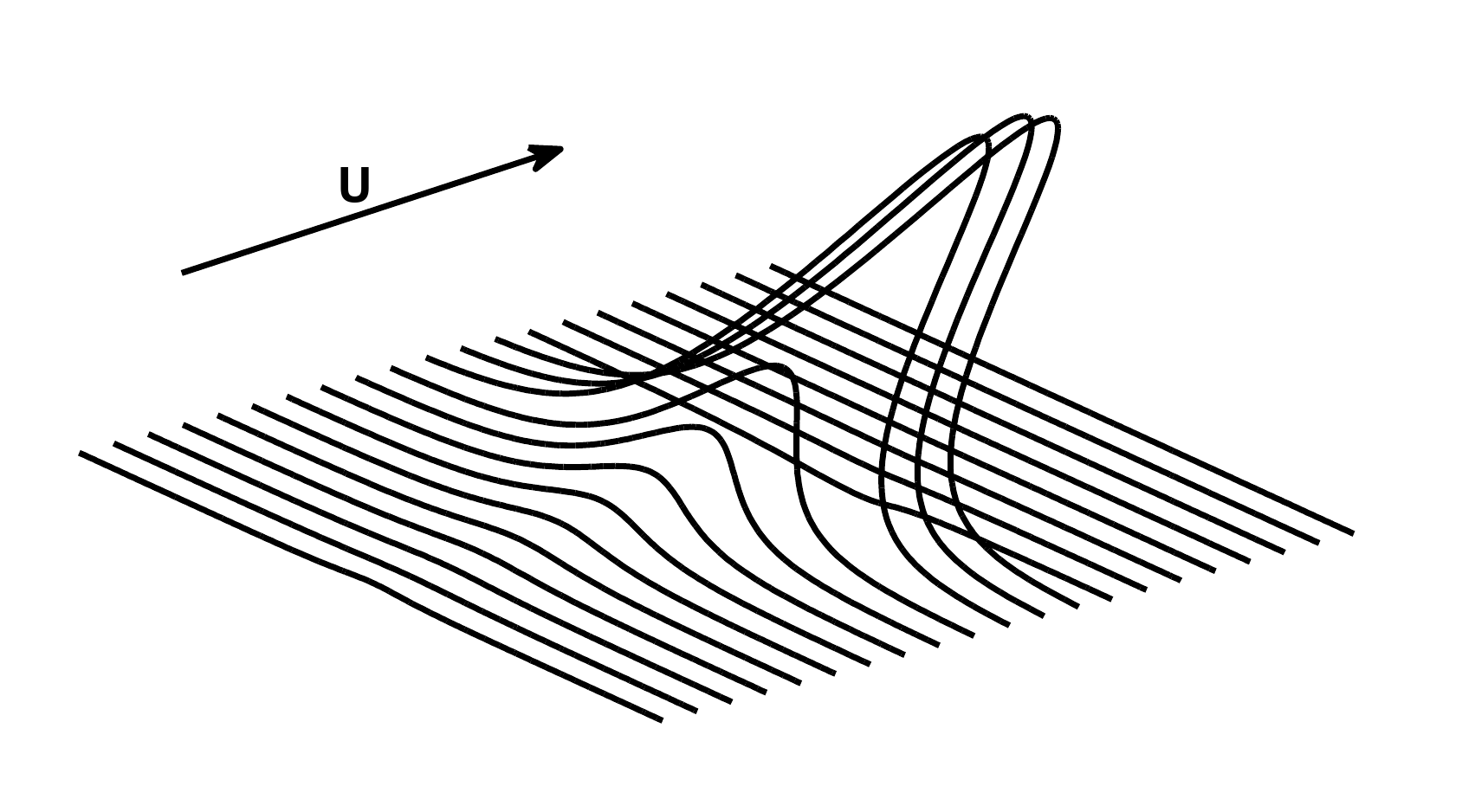}}
\caption{A horseshoe vortex bundle as originally conceived by Theodorsen (1952).}\label{hairpin}
\end{figure}

Our previous discussion has emphasized the importance of understanding  
the structure and dynamics of vorticity in pipe and channel flow. There have 
indeed been many proposals to explain and model the physics of wall-bounded
shear flows in terms of typical vortex structures.  One of the earlier attempts 
was that of Theodorsen \cite{Theodorsen52}, who suggested that the most 
significant structures are ``horseshoe vortices'' inclined at $45^\circ$ to the wall,  
which form by roll up of streamwise vorticity in the viscous sublayer.  See
Fig. \ref{hairpin}.  Considerable evidence has been found in support of 
horsehoe vortices (also known as hairpin, $\Lambda$ and $\Omega$ vortices)
from both experiment and numerical simulation 
\cite{HeadBandyopadhyay81,MoinKim85,KimMoin86,Fiedler88,Robinson91,Adrian07}. 
However, many other alternative proposals have been made. To quote from
a review of Fiedler \cite{Fiedler88}:  ``Thus, when studying the literature on 
boundary layers, one is soon lost in a zoo of structures, e.g. horseshoe- and 
hairpin-eddies, pancake- and surfboard-eddies, typical eddies, vortex rings, 
mushroom-eddies, arrowhead-eddies, etc ...''. It is fair to say that no clear 
consensus has emerged as to which vortex structures are the most relevant 
to the dynamics and statistics of the turbulence. 

Most of the effort in identifying coherent-vortices and in constructing vortex models of the 
turbulent boundary layer has gone into explaining their contribution to the Reynolds 
stress. The importance of maintaining a (nearly) constant Reynolds stress across the 
``inertial sublayer'' has been widely recognized. However, almost no effort has been made 
to explain the cross-stream transport of spanwise vorticity in terms of typical vortex structures. 
(But see \cite{Klewicki89,Klewickietal94,Priyadarshanaetal07,Klewickietal07,Bernard90}, 
discussed more below.)  Furthermore, the fundamental constraint of maintaining a constant 
mean vorticity flux across the pipe or channel has been almost universally ignored.  

Here we shall discuss such issues in the context of the ``hairpin vortex'' model of  Perry, 
Chong and collaborators \cite{PerryChong82,Perryetal86,PerryMarusic95,Marusic01,Nickelsetal07}.
This model is one of the most fully articulated attempts to explain, in analytical terms,
the features of the turbulent boundary-layer by means of representative vortex-structures. 
The Perry-Chong (PC) theory combines the Theodorsen proposal \cite{Theodorsen52} 
of ``horseshoe'' or ``hairpin'' vortices with Townsend's ``attached eddy hypothesis''
\cite{Townsend76}. In this picture, the boundary layer consists of a scale-invariant 
hierarchy of hairpin-like structures, which are attached to the wall in the sense that their 
characteristic lengths are proportional to the distance at which the eddy extends above
the wall. By averaging over an ensemble of such structures, the PC theory
yields important quantitative results, such as a logarithmic mean veocity profile 
and constant Reynolds stress in the overlap region, profiles of normal Reynolds 
stresses, tensor energy spectra as functions of distance from the wall, etc.   

PC theory is sufficiently developed that the turbulent contribution to vorticity 
flux  $\ol{\Sigma}_{yz}^{turb}=\overline{v'\omega_z'-w'\omega_y'}$ may be 
calculated by evaluating each term separately, e.g.  as integrals in wavenumber 
over the model energy spectra. However, for our purposes a simpler approach
suffices, which is to use again the identity  $\frac{\partial}{\partial y}(-\ol{u'v'})=
\overline{v'\omega_z'-w'\omega_y'}.$ We may start from the PC formula for the 
Reynolds stress as an average over hierarchy lengths $\delta$, e.g. equation (27) 
in \cite{PerryMarusic95}:
\be \ol{u_i'u_j'}=u_*^2 \int_{\delta_1}^{\Delta_E} I_{ij}\left(\frac{y}{\delta}\right) 
        p_H(\delta)\,d\delta \ee
Here $\delta_1=O(\nu/u_*)$ is the length-scale of the smallest hierarchy, whereas 
$\Delta_E=O(h)$ is the length-scale of the largest. The function $I_{ij}(y^*)$ is 
obtained by integrating over spectra of the representative vortex structures. 
See also Townsend \cite{Townsend76}, who referred to $I_{ij}(y^*)$ as an 
``eddy-intensity function.''  Both PC and Townsend required a distribution of 
hierarchy lengths $p_H(\delta)=M/\delta$ for $\delta_1<\delta<\Delta_E$ in order 
to reproduce a logarithmic mean velocity profile and a constant Reynolds stress.
We may follow their  arguments to study also the derivative of the stress,   
e.g. see \cite{PerryChong82}, Appendix B. Thus, changing integration variable
to $y^*=y/\delta$
$$ \ol{u_i'u_j'}= M u_*^2  \int_{y/\delta_1}^{y/\Delta_E} I_{ij}(y^*)\,\frac{dy^*}{y^*}. $$
and then differentiating gives
\be  \frac{d}{dy} \ol{u'v'} = \frac{M u_*^2}{y} 
             \left[I_{xy}\left(\frac{y}{\delta_1}\right)-I_{xy}\left(\frac{y}{\Delta_E}\right)\right]. 
             \lb{PC-force} \ee
Using the result $I_{xy}(y^*)\rightarrow 0$ for $y^*\gtrsim 1$ and $I_{xy}(y^*)\sim -Q y^*$ 
for $y^*\ll 1$ \cite{PerryChong82} gives
$$  \overline{v'\omega_z'-w'\omega_y'}=\frac{d}{dy}( -\ol{u'v'}) \approx 
        -\frac{MQu_*^2}{\Delta_E} $$
for $\delta_1 \lesssim y \ll \Delta_E.$ This same result may also be obtained by 
differentiating the final result for the Reynolds stress in Appendix B of
\cite{PerryChong82}. Since $\Delta_E=O(h),$ the PC theory yields a constant 
of the correct order, $\overline{v'\omega_z'-w'\omega_y'}^+=O(1/Re_*).$

These results appear satisfactory, but a little thought shows that the PC theory is 
adequate to explain vorticity flux only over the range $y_p^+<y^+\ll h^+,$ in the 
notations of the previous subsection. It is only over that range that the {\it turbulent}
vorticity flux is constant and, otherwise, the viscous diffusion is dominant. To obtain 
the correct negative sign of the turbulent flux over the range $y_p^+<y^+\ll h^+$
(and also to obtain the correct sign of the Reynolds stress \cite{PerryChong82}), 
the constant $Q$ in the above calculation must be positive. However, as we have
seen in the previous subsection, the turbulent vortex flux must reverse sign and 
become positive for $0<y^+<y^+_p.$ The PC theory, as it is presently constituted,
is inadequate to explain this sign-reversal. This is not a failure just in the buffer 
layer but over half of the range of the traditional log-layer.  

Let us consider briefly one possible remedy.
As mentioned above, it has usually 
been assumed in the  PC model that $I_{xy}(y^*)=0$ for $y^*\gtrsim 1$. For 
example, see Figure 14 of \cite{Perryetal86} or Figure 4 of \cite{PerryMarusic95}, 
where $I_{xy}(y^*)=0$ for $y^*\geq 1.$ This corresponds 
to the severe assumption that vortex hierarchy elements of length-scale $\delta$
make no contribution to Reynolds stress at wall-normal distances $y>\delta.$
If one hypothesizes  instead a slow power-law decay of the precise form 
$ I_{xy}(y^*) \sim -\frac{P}{y^*},\,\,\,\,\,y^*\gg 1, $
with $P$ another universal, positive constant, then (\ref{PC-force}) yields 
$$  \frac{d}{dy}(-\ol{u'v'}) = M u_*^2
             \left( P\frac{\delta_1}{y^2}-\frac{Q}{\Delta_E}\right) $$
for $\delta_1\ll y\ll \Delta_E,$              
in qualitative agreement with eq.(\ref{wall-scaled-force}).  The agreement 
becomes exact for a suitable choice of the constants $M,P,Q.$ An interesting
problem for future investigation might be to determine which 
``representative'' vortex structures, if any, lead to the necessary $-1$ power-law 
decay of the eddy-intensity function $I_{xy}(y^*)$ for $y^*\gg 1$. 

\subsection{Experiments and Simulations}\lb{exp-dns}

It would be very useful to get some additional clues from experiment and simulation.
There seem, however, to be very few empirical studies of the turbulent vorticity 
flux or its relevant constituent averages, $\overline{v'\omega_z'}$ and 
$\overline{w'\omega_y'}$. We know of no experimental measurements at all 
for turbulent flow in pipes or ducts. The most exhaustive investigations 
have been carried out in zero-pressure turbulent boundary layers, by Klewicki
and collaborators. Those studies include wind-tunnel boundary layers with 
$Re_\theta=1010\sim 4850$ \cite{Klewicki89,Klewickietal94} and atmospheric 
boundary layers with $Re_\theta= 2\sim 4\times 10^6$ \cite{Priyadarshanaetal07}.  
We also know of only two studies of these velocity-vorticity correlations by direct 
numerical simulation of channel flow, the $Re_*=250$ calculation of 
Leighton \& Handler reported by Bernard \cite{Bernard90} and a  $Re_*=200$ 
DNS of Crawford \& Karniadakis \cite{CrawfordKarniadakis97}. 

The picture which emerges from the lower Reynolds-number studies 
\cite{Klewicki89,Klewickietal94,Bernard90,CrawfordKarniadakis97} is as follows:
$$ \overline{v'\omega_z'}>0>\overline{w'\omega_y'} \,\,\,\,{\rm for}\,\,\,\, y^+<12$$
$$ 0>\overline{v'\omega_z'}>\overline{w'\omega_y'} \,\,\,\,{\rm for}\,\,\,\, 12<y^+<y^+_p $$
$$ 0\stackrel{?}{>} \overline{w'\omega_y'}>\overline{v'\omega_z'} \,\,\,\,{\rm for}\,\,\,\, y^+>y^+_p$$
A question mark is included in the last line since the DNS of  
\cite{Bernard90,CrawfordKarniadakis97} indicates that $\overline{w'\omega_y'}$ 
becomes slightly positive for 
$y^+\gtrsim 50>y_p^+\simeq 30.$ The results for $y^+>y^+_p$ agree rather well 
with the physical picture of vortex dynamics in the PC model \cite{PerryChong82}:
 ``the $\Lambda$-vortex stretches under its own mutual induction with its image 
 to give, in the large, plane strain
... The plane strain brings the legs of the vortex together''. 
The lifting of the negative streamwise vorticity in the head and legs 
of the hairpin produces $\overline{v'\omega_z'}<0.$ 
The simultaneous collapse of the hairpin legs towards each other, carrying
opposite signs of wall-normal vorticity, produces $\overline{w'\omega_y'}<0$.
The slight positive value of $\overline{w'\omega_y'}$ for larger $y$ values has 
been interpreted by Bernard \cite{Bernard90} also in terms of hairpin structures, 
as due to the reversal of $w'$ at some height above the counter-rotating legs of 
the vortex.   Our calculation in the previous subsection indicates that 
$\overline{w'\omega_y'}>\overline{v'\omega_z'},$ yielding a net negative vorticity 
flux.  For more detailed discussion of these correlations in terms of typical vortex 
structures, see \cite{Klewickietal94}, Section III and \cite{Bernard90}, Section 4. 

 On the other hand,  the PC model does not seem to accord well with the 
lower Reynolds observations for $y^+<y^+_p.$ Klewicki et al. \cite{Klewickietal94} 
have made a detailed quadrant analysis of the contributions to $\overline{v'\omega_z'}$
at $y^+=5.3, 14.2$ and $26.3$ in the BL at $Re_\theta=1010$ with $y_p^+
\doteq 40.$ Although $\overline{v'\omega_z'}<0$ for  $y^+=14.2$ and $26.3,$
as the hairpin picture might suggest, \cite{Klewickietal94} have found that 
the dominant events are 2nd-quadrant, i.e.  $v'<0$ and $\omega_z'>0.$
Furthermore there is strong dependence on Reynolds number,  with the 
magnitude of $\overline{v'\omega_z'}$ decreasing sharply with increasing 
Reynolds number for $Re_\theta=1010\sim 4850.$



These tendencies persist in the high $Re_\theta$ data from the ABL in 
\cite{Priyadarshanaetal07}. In fact, their smooth-wall data at $Re_\theta=2\times 10^6$
show $\overline{v'\omega_z'}>0$ for the whole range of $y^+.$ Furthermore,  
$\overline{w'\omega_y'}$ appears to change sign, being negative for 
$y^+<y^+_p$ and positive for $y^+>y^+_p$ (using rough-wall data). 
The sign for $y^+>y^+_p$ may be due to growth of hairpin vortices 
in the spanwise direction \cite{Adrian07,Adrianetal01,TomkinsAdrian03}.  
Clearly, further experimental and numerical studies would be most helpful,
in order to verify the limited data and to map out all the dependencies on 
wall-distance, Reynolds number, roughness, etc. Particularly illuminating 
would be conditional sampling studies, with VISA (variable-interval 
space-averaging) sampling technique and quadrant conditions 
like those applied to Reynolds stress \cite{MoinKim85,KimMoin86}, but now 
carried out for the vorticity flux. Such investigations could help to reveal the vortex 
structures and dynamics which contribute most significantly to the cross-stream 
transport of spanwise vorticity. 
               

\section{Comparison of Classical and Quantum Fluids}\lb{superfluid}

We shall now discuss briefly some of the corresponding problems in the field
of quantum superfluids.
A complete review of superfluidity and superconductivity would be out of place
here and we refer the reader interested in further background to standard texts
\cite{Tinkham96, Annett04}. We just remark on a few basic facts, the most 
important of which is quantization. As first noted by Onsager \cite{Onsager49} 
and Feynman \cite{Feynman55}, the vorticity in neutral superfluids resides 
entirely in thin filaments whose circulation is quantized in units of 
$\kappa=h/m,$ where $h$ is Planck's constant and $m$ is the atomic mass 
of the condensed Bose particle. For liquid ${\,\!}^4{\rm He},$ $\kappa\approx 9.97
\times 10^{-4}\,\,{\rm cm}^2/{\rm sec}$ and the core radius of the vortex lines 
is of order of angstroms. An analogous fact holds for charged superfluids 
or superconductors, in which the condensed Bose ``particle'' is generally a 
Cooper pair of two electrons of charge $2e.$ As first pointed out by London
\cite{London48}, the magnetic flux in a superconductor is likewise quantized 
in units of $\Phi_0=h/2e=2.07\times 10^{-15}$ webers. London considered 
what are now called ``Type-I superconductors'' which are perfect diamagnets 
and expel all magnetic fields (the Meissner effect). Thus, he had in mind the 
magnetic flux through superconducting rings and other non-simply-connected
samples. However, it was pointed out later by Abrikosov \cite{Abrikosov52,Abrikosov57}
that there is another class of ``Type-II superconductors'' which are penetrated 
by magnetic flux tubes with flux quantized in units of $\Phi_0.$

We now discuss the ``phase-slip'' phenomenon, which is closely analogous 
to the results for classical turbulence presented in section \ref{vort-flux-press-grad}. 
These ideas were first employed by Josephson for superconductors and 
extended further for both superfluids and superconductors by Anderson, hence
often called the ``Josephson-Anderson relation'' \cite{Packard98}. The classic
presentation is in the review article by Anderson \cite{Anderson66}. He argued 
there that the drop in chemical potential $\mu$ (per unit mass) between two 
points $P_1$ and $P_2$ in a neutral superfluid should be related to the rate 
$dn/dt$ at which quantized vortex-lines cross an oriented line between the
two points:
\be \ol{\mu(P_2)}-\ol{\mu(P_1)}=\kappa \,\ol{dn/dt } \lb{chem-drop} \ee
Here the overbar indicates a time-average. In an appendix---entitled ``A `New'
Corollary in Classical Hydrodynamics?"---Anderson also pointed out that 
this relation for quantum fluids has a precise classical analogue, of the form:
\be \ol{(p+U+\frac{1}{2}|\bu|^2)}_{P_2} - \ol{(p+U+\frac{1}{2}|\bu|^2)}_{P_1}
                  = \int_{P_1}^{P_2} \ol{\bu\btimes\bomega}\bdot d\bx. 
                  \lb{pressure-drop} \ee
This is just the relation (\ref{classical-JA}) that we derived earlier, in the special 
case of an inviscid fluid, with no body-forces, and in line-integral form. It is a
generalization of the relation noted by Taylor \cite{Taylor32} for shear turbulence.  
A similar relation can be obtained for type-II superconductors
\be \ol{V(P_2)}-\ol{V(P_1)}=\Phi_0 \,\ol{dn/dt }, \lb{volt-drop} \ee
now relating the difference in voltage between two points to the rate at which
a line from $P_1$ to $P_2$ is crossed by quantized flux \cite{Anderson62,Josephson65}. 
This relation was the basis of Anderson's ``flux-creep theory'' to explain resistive 
behavior in type-II superconductors above a critical current. 

These ideas were perhaps most carefully elaborated by Huggins \cite{Huggins70}.
His paper is especially interesting for us, because he employed the classical 
fluid equations as his model, despite intended applications to quantum superfluids. 
This is legitimate, because a superfluid will obey the laws of classical hydrodynamics
except close to the cores of vortex lines. Huggins makes several assumptions that
are adapted to the superfluid problem, e.g. that all the vorticity is confined to 
compact vortex tubes with total circulation of magnitude $\kappa.$ 
However, almost all of his discussion can  be easily generalized to the case of continuous 
vorticity distributions of any magnitude (see Appendix \ref{detailedJ}). 

The starting point of Huggin's treatment is Kelvin's minimum kinetic energy theorem 
\cite{Kelvin49}. As is well-known, Kelvin showed that, among all incompressible 
velocity fields with the same boundary conditions in a finite domain $\Omega,$
the one with minimum energy is the potential flow. To prove this, Kelvin decomposed
the velocity field as $\bu=\bu_\phi+\bu_\omega$ where $\bu_\phi=\grad \phi$ is 
the potential flow with the specified inflow-outflow boundary conditions and 
$\bu_\omega$ is a field containing all of the vorticity of the fluid but with no flow
through the boundary. In that case the kinetic energy decomposes into a sum 
of kinetic energies for the potential and vortex fields, so that $E=E_\phi+E_\omega
\geq E_\phi.$ Huggins discusses a classical fluid model in the form
\be \partial_t\bu = -\grad h + \bu\btimes\bomega + \bg, \lb{GNS} \ee
where the enthalpy $h=p+U+\frac{1}{2}|\bu|^2$ includes the contribution of 
any potential energy $U$ whereas $\bg$ contains all the non-potential forces,  
e.g. $\bg=-\nu(\grad\btimes\bomega)+\bF$ if there is only a viscous force and a 
non-conservative body-force.  Huggins further considers flow 
through a channel driven by a difference in the enthalpy 
$h_{in}$ at the inflow section and enthalpy $h_{out}$ at the outflow 
section. In this setting \cite{Huggins70} proves that
\be \dot{E} = J(h_{in}-h_{out}) + \int_\Omega \varrho\bu\bdot\bg \,d^3x \lb{huggins-all} \ee
\be \dot{E}_\phi = J(h_{in}-h_{out}) -\kappa \dot{J}_\phi \lb{huggins-phi} \ee
\be \dot{E}_\omega = \kappa \dot{J}_\phi + \int_\Omega \varrho\bu\bdot\bg \,d^3x 
\lb{huggins-vort} \ee
Let us consider each of these relations in turn.

The first result (\ref{huggins-all}) is just the standard energy balance relation. Since 
$J$ is the total mass flux through the pipe, $J(h_{in}-h_{out}) $ equals the energy input 
into the fluid through work by the applied enthalpy difference. On the other hand, 
$D=-\int_\Omega \varrho\bu\bdot\bg \,d^3x$ just represents the dissipation by the 
non-potential forces. The second relation (\ref{huggins-phi}) is more interesting.
This result shows that the energy supplied by the enthalpy difference goes entirely
into the potential flow. What acts as ``dissipation'' on the potential flow is the term
$\kappa \dot{J}_\phi,$ which transfers energy to the vortex system. As seen from 
the final relation (\ref{huggins-vort}), this energy is then dissipated by the non-potential 
forces, e.g. ordinary viscosity, acting on the vortices. The transfer term is the product
of the quantum of circulation $\kappa$ and the rate $\dot{J}_\phi$ at which vortex
lines are crossing the mass flux in the potential flow. For example, for a single 
length of vortex line    
\be \kappa \dot{J}_\phi =  \kappa \int_{vort} \varrho\bu_\phi\bdot \bu_v\btimes d{\bf l} 
\lb{mass-cross} \ee
The quantity $\bu_v$ is the velocity of the element $d{\bf l}$ of vortex line,
so that $\bu_v\btimes d{\bf l}$ represents the rate at which the line-element 
sweeps out area. It is easy to recover Anderson's relation (\ref{chem-drop}) 
[or (\ref{pressure-drop})] from (\ref{huggins-phi}). For example, if vortex 
lines are crossing the entire mass flux $J$ in the channel at a frequency $\nu,$
then $\dot{J}_\phi=J\nu.$ In that case, $h_{in}-h_{out}=\kappa\nu$ for steady-state flow. 

Note that it is crucial in (\ref{mass-cross}) that $\bu_v\neq \bu_\phi,$ or otherwise
$\dot{J}_\phi=0.$ Thus, it is important that quantized vortex lines are {\it not} material 
lines, moving with the local potential velocity $\bu_\phi.$ An analogous fact is 
true for vortex lines in classical flow which, because of the non-potential force
$\bg,$ do not move with the local fluid velocity $\bu$. In general one may 
decompose the non-potential force into components transverse and longitudinal 
to the local vorticity vector: 
\be  \bg=\Delta\bu\btimes \bomega+\alpha\bomega, \lb{decomp} \ee
where $\Delta \bu=\bomega\btimes\bg/|\bomega|^2,\,\,
\alpha=\bomega\bdot\bg/|\bomega|^2.$ The vector $\Delta\bu$ can be interpreted
as a ``drift velocity'' of vortex lines with respect to the local fluid velocity, so that 
the vortex line moves with net velocity $\bu_v=\bu+\Delta \bu.$ In that case, 
the transverse force has the form of a ``Magnus force'' $(\bu_v-\bu)\btimes \bomega,$
similar to that which acts on superfluid vortices \cite{Thoulessetal99,AoZhu99}. 
On the other hand, there are non-vanishing longitudinal forces as well in classical 
fluids and these lead to non-trivial effects, such as helicity cascade 
\cite{Brissaudetal73} and kinematic $\alpha$-effect \cite{Frischetal87}.
The effect of both terms on vortex motion is clearly seen in turbulent 
flows, where viscosity leads to quite different properties of material lines 
and vortex lines \cite{Gualaetal05}. These effects appear to persist even 
in the zero-viscosity limit, when vortex lines behave as material lines only in 
an average sense \cite{Eyink06b,Chenetal06}. 

It is interesting to note that the phase-slips of quantized flux lines and their
consequent dissipation lead to a ``drag reduction problem'' in superconductor
technology. In type-II superconductors the result analogous to Huggins' above 
is that crossing of supercurrent by flux lines at a rate $\dot{I}_\phi$ produces 
an energy dissipation $I(V_1-V_2)=\Phi_0\dot{I}_\phi,$ where $I$ is the total 
electric current. This motion of vortices produces a ``quasi-resistance'' in 
type-II superconductors above some critical current $I_c,$ which destroys 
their useful superconducting properties. This problem is particularly 
severe for the newer class of high-temperature superconductors (HTS), 
since the vortices there are subject to weaker restoring forces and stronger
thermal fluctuations. Unlike the vortices in conventional type-II superconductors 
which are frozen into an Abrikosov lattice or vortex crystal, the vortices in HTS 
melt at high enough temperatures into a mobile ``vortex liquid'' that produces 
sizable resistivity. See \cite{Blatteretal94,Norton98,Gurevich07} and, for an 
excellent popular account, \cite{Bishopetal93}.  One solution to this problem 
is to ``pin'' the vortices to prevent their motion. Methods that have proved successful 
in more conventional type-II superconductors are to dope the superconductor 
with impurities or to manufacture wires and films from sintered powders 
(e.g. see \cite{Matsumoto06,Senkowiczetal08} as representative papers). 
Anything that prevents free cross-current motion of the quantized vortices 
restores the superconducting properties and permits electric current to 
flow with no resistance. 
 
An even closer analogy exists between classical turbulence and turbulence
in neutral superfluids, as originally conceived by Feynman \cite{Feynman55}.
To quote his own words:
\begin{quote}
``In ordinary fluids flowing rapidly and with very low viscosity the phenomenon 
of turbulence sets in. A motion involving vorticity is unstable. The vortex 
lines twist about in an ever more complex fashion, increasing their length 
at the expense of the kinetic energy of the main stream. That is, if a 
liquid is flowing at a uniform velocity and a vortex line is started somewhere 
upstream, this line is twisted into a long complex tangle further downstream. 
To the uniform velocity is added a complex irregular velocity field. The 
energy for this is supplied by pressure head. We may imagine that similar
things happen in the helium.''$\,\,\,\,\,\,\,\,\,$ ---Feynman (1955)
\end{quote}
Experiments on turbulent flow of superfluid ${\,\!}^4{\rm He}$ in circular
pipes  \cite{Swansonetal00} and rectangular ducts \cite{XuSciver07} 
show similar behaviors as classical turbulence, both in drag resistence 
as functions of Reynolds number and in mean velocity profiles. In 
addition to the complex, random motion of vortex lines discussed by
Feynman above, there must also be organized motion. Just as in 
the classical case, there must be a ``cascade'' of azimuthal vorticity 
along the radial direction toward the pipe axis. As follows from our discussion 
above, this is essential for the vortex lines to extract energy from the 
pressure head. Vortex structures like ``hairpins,'' ``rings,''  etc. will also 
doubtless form and contribute to the necessary cross-stream flux of vorticity. 

\section{Discussion and Conclusions}\lb{discussion}

We have presented in this paper a systematic picture of pressure-driven
turbulence in pipes and channels as spatial ``inverse cascades'' of spanwise
vorticity in the cross-stream direction. This complements the more conventional 
picture of these flows as ``forward cascades'' of streamwise momentum,  
also across the stream but toward the viscous layer at the wall. Following 
Huggins \cite{Huggins94}, we have shown that the spatial flux of vorticity 
is necessary to produce a pressure-drop down the pipe and consequent 
dissipation.  This flux is dominated by viscous diffusion of vorticity at 
distances to the wall less than the peak of the Reynolds stress, 
well into the classical log-layer. The Perry-Chong model based 
on  ``representative'' hairpin/horsehoe vortices has been 
shown to be inadequate to describe the cross-stream vorticity flux over 
that range, but viable at distances greater than the Reynolds-stress 
maximum. We have shown that similar results hold for quantum 
superconductors and  superfluids, and, especially, for superfluid 
turbulence driven by applied pressure gradients in pipes and channels.  

We hope that one consequence of this work will be to stimulate a greater 
effort by experimentalists and numericists to discover the detailed 
vortex structures and dynamics which contribute to the cross-stream 
vorticity transport, as a function of distance to the wall and of Reynolds 
number. Theorists should also recognize the condition of constant 
wall-normal vorticity flux as an important constraint on statistical models
and closures. On the practical side, we cannot emphasize too strongly
that {\it pressure-drop and cross-stream transport of spanwise vorticity
are mathematically equivalent.} This is a very universal
fluid-dynamical result, applicable not only to turbulence but also to 
laminar and transitional flow. See eqs.~(\ref{huggins-phi}),(\ref{huggins-vort}) 
or the very general formulation in eq.~(\ref{dJA}) of Appendix B, for instance.
In fully-developed turbulence in channels and pipes, which enjoys all the statistical 
symmetries permitted by the boundary conditions, this dynamical constraint 
leads to constant time-average flux of spanwise vorticity across the main flow. 
The laminar solution in channel geometry with parabolic velocity profile has 
also a constant vorticity-flux, given entirely by viscous transfer (see 
\cite{Huggins94}, section 13). In the turbulent case, there is an 
additional mean contribution from nonlinear transport. For a fixed mass-flux, 
the vorticity transfer in turbulent flow is much greater than in the laminar state,
leading to a greatly enhanced streamwise pressure drop and enhanced energy
dissipation. Any method for reducing drag must therefore reduce the  cross-stream
transfer of vorticity. It is hoped that the vortex-cascade framework presented here 
will provide new perspectives on the difficult problems of wall-bounded turbulence.

A fresh view of the role of the vorticity may be of use, in particular, in 
the problem of polymer drag reduction. It has long been suspected that the 
critical action of polymers in turbulent drag reduction is through their interaction 
with vortices \cite{Dubiefetal05,Kimetal08,Yarin97}. The same is also true for various 
other drag-control strategies, such as with electromagnetic body forces 
\cite{CrawfordKarniadakis97}. A persistent puzzle has been to understand
how polymer-induced changes in vortex structure are mediated into 
substantial changes of the large-scale structure of the flow. The vorticity
flux tensor (\ref{Sigma-tensor}) is a natural locus for critical polymer activity.
There is generally a scale separation in high Reynolds number turbulence 
between the peaks of the spectra of the velocity and the vorticity. However, 
\cite{Priyadarshanaetal07} have found that the velocity-vorticity cospectra 
in boundary layers  are not generally dominated by intermediate wavenumbers 
between these two peaks, but instead get dominant contributions from 
wavenumbers near the peak in the associated vorticity spectra and, 
in some cases, also near the peak in the associated velocity spectra. 
Thus the velocity-vorticity correlations that contribute to the turbulent
vorticity flux are a natural bridge between the small-scale and large-scale 
features of the flow.  

\vspace{.3in} 

\noindent {\small
{\bf Acknowledgements.} I thank C. Barenghi, P. Bernard, S. Chen, P. Hohenberg, 
J. Klewicki,  A. Leonard, C. Meneveau, S. Pope, K.  Sreenivasan and J. Wallace 
for discussions. I am particularly grateful to P. Ao for suggesting a connection 
between turbulent dissipation  and the Josephson-Anderson relation in superfluids. 
I also thank R. J. Adrian for helping to stimulate my early interest in turbulent boundary 
layers and their vortical structures. This work was supported by NSF grant 
\# ASE-0428325 at the Johns Hopkins University and by the Center for 
Nonlinear Studies at Los Alamos National Laboratory.}

\newpage

\setcounter{section}{0}
\renewcommand{\thesection}{\Alph{section}}

\section{Appendix: The Vorticity Source Density at a Wall} 

It is the purpose of this appendix to derive the following expression 
$$\bsigma = \hn' \times [ \nu \grad \times \bomega-\bF]$$
for the vorticity source density at the wall. Here, in contrast to the main 
text, we have preferred  to use the {\it inward}-pointing unit normal 
$\hn'= - \hn$ at the boundary $\partial\Omega$ of the domain.
Let us consider any unit vector ${\bf e}_i$ locally tangent to the surface, 
$\hn \cdot {\bf e}_i = 0$, and consider the small square loop $C$ normal 
to ${\bf e}_i$, with one edge lying in the boundary surface $\partial \Omega$:
\begin{figure}[h!]
\centerline{\includegraphics[width=350pt,height=200pt]{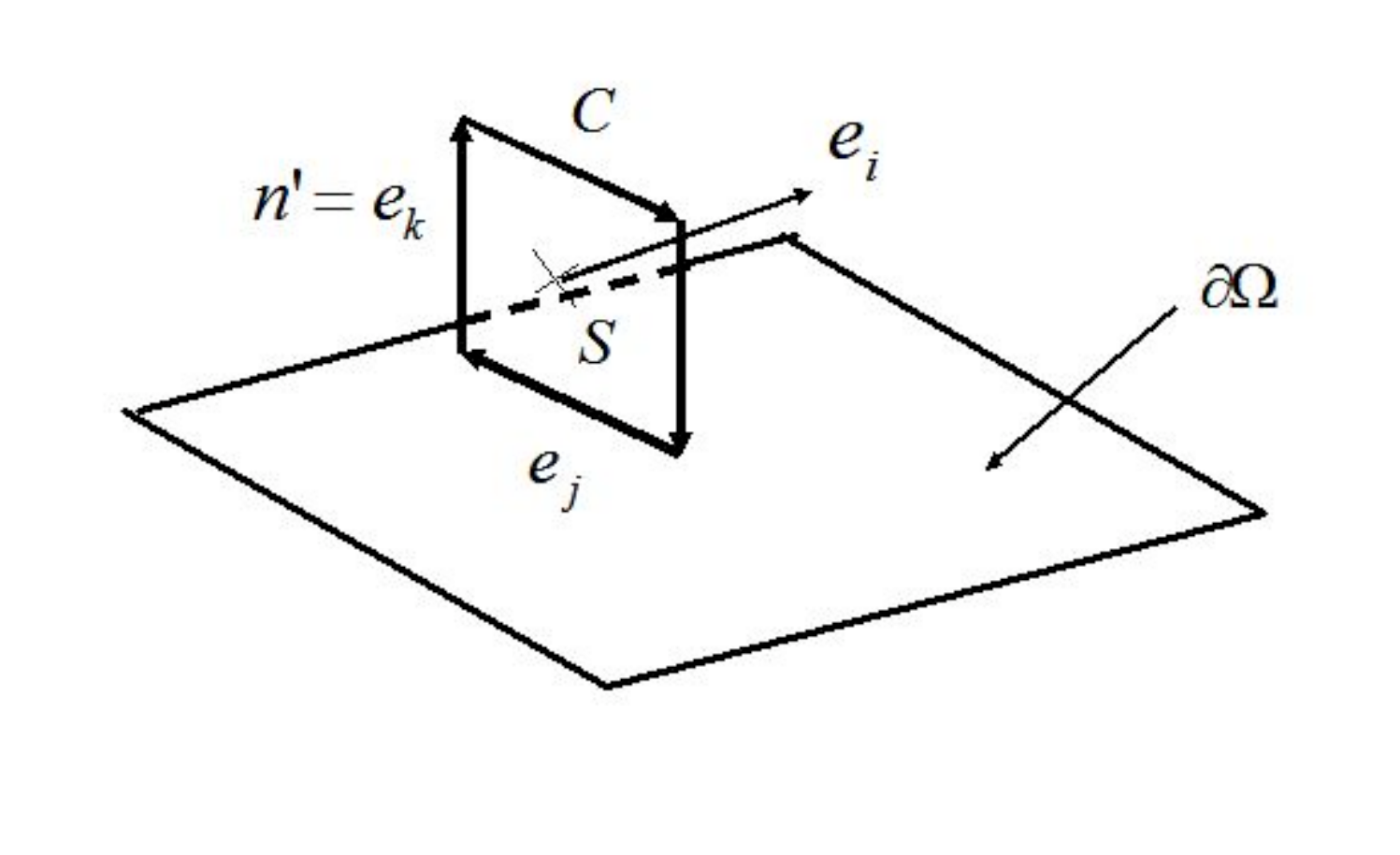}}
\caption{A test loop orthogonal to the wall.}\label{fig:loop}
\end{figure}

\noindent Here, ${\bf e}_i \times {\bf e}_j = {\bf e}_k = \hn'$ and the edge 
tangent to ${\bf e}_j$ lies in the surface $\partial \Omega$.
If $S$ is the square surface bounded by $C,$
$$ \int_S \omega_i dx_j dx_k = \oint_C \bu \cdot d\bx. $$
Letting $S(t)$ and $C(t)$ be the surface $S$ and loop $C$ materially 
advected by the fluid, then by the Kelvin Theorem
$$\frac{d}{dt} \int_{S(t)} \omega_i dx_j dx_k |_{t=0} = \frac{d}{dt} 
\oint_{C(t)} \bu(t) \cdot d \bx|_{t=0} = 
\oint_C [-\nu\grad \times \bomega+\bF] \cdot d \bx. $$
The bottom segment parallel to ${\bf e}_j$---which is stationary 
in the surface because of the stick conditions---contributes a term
$$ \int_{C \cap \partial \Omega} [-\nu\grad \times \bomega+\bF]_j d x_j$$
to the time-derivative of the flux $\int_S \omega_i dA$. We therefore identify
$$ \sigma_i = - \nu (\grad \times \bomega)_j +f_j $$
or, for any direction,
$$ \bsigma = \hn' \times [\nu \grad \times \bomega-\bF], $$
in agreement with our earlier finding.

The above argument also reveals the precise meaning of the ``vorticity 
source density'' $\bsigma$. If one considers any loop $C$ with part 
of its length in the boundary $\partial \Omega$, then the contribution of the boundary 
segment to the time-derivative of the vorticity flux through $C$ is given by
$$ \int_{C \cap \partial \Omega} \bsigma \cdot (\hT \times \hn') d s $$
where $\hT$ is the unit tangent vector along the curve $C$, for a specified 
orientation. With this convention the flux is measured positive in the direction 
to the righthand side as one traverses the curve $C$, i.e. parallel to $\hT \times \hn'$.

The expression that we have derived for the ``vorticity source density''
$$ \bsigma = \hn' \times [\nu \grad \times \bomega-\bF] $$
is the same as that obtained by Lyman \cite{Lyman90}. There seems, however, to 
be some controversy about this result. It has been argued by Wu \& Wu 
\cite{WuWu96} that Lyman's prescription (and ours!) is incorrect and they suggest 
that $\hn' \times \nu (\grad \times \bomega)$ be replaced instead with 
$$ {\bsigma}_{\mbox{Wu}} = \nu \frac{\partial \bomega}{\partial n} 
= \nu (\hn \cdot \grad) \bomega $$
with outward-pointing normal $\hn$. The two expressions have equal integrals 
over the boundary $\partial\Omega$ of the flow-domain since, by the divergence 
theorem and vector-calculus identities, they both equal $\int_\Omega \nu\nabla^2
\bomega\,\,d^3x.$ The two differ locally by a term $(\grad\hn)\bomega,$ which 
vanishes for a flat surface. In particular, the two expressions agree for channel flow. 
However, with the precise interpretation of the vorticity source density given by us
above, our result is a direct consequence of the Kelvin Theorem. 
We note furthermore that $\bsigma$ ought to lie parallel to vorticity vectors generated 
within the boundary surface and thus satisfy $\hn \bdot \bsigma =0$. While Lyman's 
prescription always satisfies this condition, that of Wu \& Wu does not in general. 

\section{Appendix: The Detailed Josephson Relation}\lb{detailedJ}

We present here a proof of the ``detailed Josephson relation,'' due to 
Huggins \cite{Huggins70}. Although all of the results presented below are 
essentially contained in his paper, the generality of those results is somewhat 
obscured by various special assumptions invoked in that work. Our derivation 
has a slight novelty and hopefully makes the main ideas more clear.  

We consider a generalized channel geometry, where the simply-connected
flow domain $\Omega$ has boundary consisting of an in-flow surface $S_{in},$
an out-flow surface $S_{out},$ and wall surface $S_w.$ The boundary 
conditions on the velocity $\bu$ are:
\be  \left.\bu\right|_{S_{in}}=\bu_{in},\,\,\,\,
     \left.\bu\right|_{S_{out}}=\bu_{out},\,\,\,\,
     \left.\bu\right|_{S_{w}}=\bzed. 
 \lb{vel-bc} \ee
Huggin's argument builds upon the proof of the Kelvin minimum-energy 
theorem \cite{Kelvin49}. Thus, the velocity is decomposed as $\bu=\bu_\phi+\bu_\omega,$
where $\bu_\phi=\grad\phi$ is the unique potential flow field satisfying the incompressibility 
constraint $\grad\bdot\bu_\phi=\bigtriangleup\phi=0$ and the through-flow 
boundary conditions  
\be  \left.\hn\bdot\bu_\phi\right|_{S_{in}}=\hn\bdot\bu_{in},\,\,\,\,
     \left.\hn\bdot\bu_\phi\right|_{S_{out}}=\hn\bdot\bu_{out},\,\,\,\,
     \left.\hn\bdot\bu_\phi\right|_{S_{w}}=\bzed,  
 \lb{pot-vel-bc} \ee
inherited from the full velocity $\bu.$ Thus, $\bu_\omega=\bu-\bu_\phi$ 
is also incompressible, carries the full vorticity of the flow ($\bomega=
\grad\btimes\bu_\omega$), and satisfies the no-through b.c. 
$\left.\hn\bdot\bu_\omega\right|_{\partial\Omega}=0.$ It follows from these facts that
$$ \int_\Omega \bu_\phi\bdot\bu_\omega \,\,d^3x
     =\int_\Omega \grad\bdot[\phi\bu_\omega] \,\,d^3x
     =\int_{\partial\Omega} \phi\,\bu_\omega\bdot\hn\,\,dA=0. $$ 
This orthogonality of $\bu_\phi$ and $\bu_\omega$ is, of course, the essence of 
Kelvin's theorem. 

One can now substitute $\dot{\bu}_\phi=\grad\dot{\phi}$ into the equation of motion
(\ref{GNS}) to obtain an equation for the time-derivative of $\bu_\omega:$
$$\dot{\bu}_\omega = -\grad h' +\bu\btimes \bomega +\bg$$
with $h'=h+\dot{\phi}=p+U+\frac{1}{2}|\bu|^2+\dot{\phi}.$ This result may 
integrated along any path $C$ in the fluid from point $P_1$ to $P_2,$ 
yielding 
\be  h'(P_2)-h'(P_1)+ \int_{P_1}^{P_2} \dot{\bu}_\omega\bdot d\bx
       = \int_{P_1}^{P_2} (\bu\btimes \bomega +\bg)\bdot d\bx. \lb{almost} \ee
This result is already close to the desired relation. For example, if one 
Reynolds-averages with respect to a stationary ensemble, then 
$ \ol{\dot{\bu}_\omega}=\bzed$ and thus (\ref{almost}) reduces to 
Anderson's relation (\ref{pressure-drop}), in a slightly generalized form. 

The key observation of Huggins is that another kind of average, a ``mass-flux
average,'' also makes the time-derivative term vanish. To explain this average,
we must say a few words about  ``intrinsic'' or ``natural'' coordinates for  the 
incompressible potential flow $\bu_\phi.$  This system takes 
as coordinate surfaces the iso-potential surfaces $S_\phi$ and the streamlines of 
$\bu_\phi.$ As a coordinate variable along streamlines one may take  $\phi$ itself 
or alternatively arclength $\ell,$  related by $d\phi=u_\phi d\ell.$ The line-vector
element along streamlines is thus $d\bell=\widehat{\bu}_\phi\, d\ell.$ On 
a selected iso-potential surface $S_\phi$ one may introduce curvilinear 
coordinates $\psi,\chi$ to label the individual streamlines, completing
the set $(\phi,\psi,\chi)$ of natural/intrinsic coordinates. The area element 
on $S_\phi$ is given by $dA=|\frac{\partial \bx}{\partial \psi}\btimes
\frac{\partial\bx}{\partial\chi}| d\psi d\chi,$ in terms 
of which another positive measure may be defined as $dJ=
\varrho u_\phi dA.$ This measure represents the mass-flux carried by the 
potential flow $\bu_\phi$ across each subset of the surface $S_\phi$. Note 
that it does not matter in the definition of $dJ$ which iso-potential surface 
is considered, since $dJ$ is constant along streamlines by incompressibility.   
For a general surface transverse to the potential flow, not necessarily an 
equi-potential surface, $dJ=\varrho|\bu_\phi \bdot\hn| dA.$ The entire 
construction simplifies in two space dimensions.  In 2D intrinsic orthogonal 
coordinates are $(\phi,\psi),$ where $\psi$ is the streamfunction which is the 
harmonic conjugate to the potential $\phi.$ In that case $dJ=\varrho d\psi.$ 

The derivation is now easily completed. Take the integral in (\ref{almost}) along the 
streamlines crossing from surface $S_{in}$ to surface $S_{out},$ and average
with respect to $dJ.$ It then follows from the b.c. (\ref{pot-vel-bc}) that the mass-flux
of the potential flow is equal to the mass-flux of the full flow. The key term to analyze 
is $\int dJ\int \dot{\bu}_\omega\bdot d\bell.$ Using 
$\varrho \bu_\phi \, d^3 x = \varrho u_\phi dA \cdot \widehat{\bu}_\phi d\ell
      = dJ\, d\bell, $
this may be rewritten as 
$$\int dJ \int \dot{\bu}_\omega\bdot d\bell
                              =\varrho\int \dot{\bu}_\omega\bdot \bu_\phi\, d^3 x=0.$$
The latter integral vanishes by the same reasoning used to prove 
orthogonality of $\bu_\phi$ and $\bu_\omega.$ We finally obtain the 
{\it detailed Josephson relation} of \cite{Huggins70}:
\be \int dJ \,h_{in}' - \int dJ \,h_{out}' =   
                               -\int dJ \int (\bu\btimes \bomega +\bg) \bdot d\bell \lb{dJA} \ee                            
The line-integrals on the righthand side are taken along streamlines of the 
potential flow and represent the generation of circulation along those lines. 
Substituting the decomposition (\ref{decomp}) for $\bg$ one can see 
that there are two essential contributions, one from the line-integral of 
$\bu_v\btimes\bomega$ which represents the transverse motion of 
vorticity across streamlines at velocity $\bu_v=\bu+\Delta\bu$ and the
other from the line-integral of $\alpha\bomega$ which represents 
acceleration by longitudinal forces. Note that one may also write
$$ \int (\bu\btimes \bomega +\bg) \bdot d\bell = \frac{1}{2}\int
                        \epsilon_{ijk}\Sigma_{ij} d\ell_k, $$
in terms of the vorticity flux tensor (\ref{Sigma}). The lefthand side of 
(\ref{dJA}) may be rewritten as
$$  \int dJ \,h_{in}' - \int dJ \,h_{out}'  = J \left[ \langle h_{in}'\rangle_J - \langle
h_{out}' \rangle_J \right],$$
where $J$ is the total mass flux and $\langle\cdot\rangle_J=
(1/J)\int (\cdot) dJ$ defines the mass-flux average. The energy relations 
(\ref{huggins-all})-(\ref{huggins-vort}) may be similarly generalized, with 
$J (h_{in}- h_{out})$ in (\ref{huggins-all}) and (\ref{huggins-phi}) replaced with  
$J \left[ \langle h_{in}\rangle_J - \langle h_{out} \rangle_J \right],$ and the
transfer term $\kappa\dot{J}_\phi$ in (\ref{huggins-phi}) and (\ref{huggins-vort})
replaced with the righthand side of (\ref{dJA}). For details of the proofs, 
see \cite{Huggins70}.



\begin{thebibliography}{99}
\bibitem{footnote1}
Except possibly at isolated points where $\bsigma=\bzed,$ in which case a normal 
component is permitted.
\bibitem{footnote2}
The pressure conditions,  
$\hn \btimes \grad p =-\hn\btimes (D_t \bu+\nu\grad\btimes\bomega-\bF)$, which arises from the 
tangential components of the equation of motion at the boundary, are not automatically 
satisfied in Leray's existence proof of Navier-Stokes solutions, which instead enforces the 
Neumann condition  $\hn \bdot \grad p =-\hn\bdot (D_t \bu+\nu\grad\btimes\bomega-\bF)$.
However, Temam \cite{Temam82,Temam06} has proved that the tangential components
of the equation of motion at the boundary emerge as necessary and sufficient ``compatibility
conditions'' for a solution to remain smooth up to the wall.
\end{thebibliography}
\end{document}